\documentstyle[12pt]{article}
\begin{document}

\begin{flushright}
September 1999

OU-HET 325
\end{flushright}

\begin{center}

\vspace{5cm}
{\Large Monopoles and Black Hole Entropy}

\vspace{2cm}
Takao Suyama \footnote{e-mail address : suyama@funpth.phys.sci.osaka-u.ac.jp}

\vspace{1cm}

{\it Department of Physics, Graduate School of Science, Osaka University, }

{\it Toyonaka, Osaka, 560-0043, Japan}

\vspace{4cm}

{\bf Abstract} 

\end{center}

We consider the entropy of a black hole which has zero area horizon.
The microstates appear as monopole solutions of the effective theory on the corresponding 
brane configurations.
The resulting entropy formula coincides with the one expected from stringy calculation 
and agrees with U-duality invariance of entropy.

\newpage

{\large {\bf 1.\ Introduction}}

\medskip

Black holes have been found to be good objects to examine the power of string theory.
The entropy of a 5 dimensional extremal black hole was derived by using the D-brane 
description \cite{SV}\cite{CM}.
It agrees with Bekenstein-Hawking entropy \cite{BH} calculated from the corresponding 
black hole solution in supergravity \cite{solution}.
It was also argued that such D-brane description could apply to non-extremal black holes and 
their Hawking radiation etc was discussed \cite{CM}.
This technology was applied to 4 dimensional black holes, and the entropy formula was 
derived for extremal cases \cite{4dim1}\cite{4dim2}.

It is known that 4 dimensional extremal black holes need to have 4 charges to have 
finite area horizon.
For example, compactifying D1-D5 system considered in \cite{CM} on $S^1$ provides a 
4 dimensional black hole, but its horizon area is found to be zero.
This seems to indicate that the entropy is zero.
However, the method to count microstates used in \cite{CM} is independent of whether 
spatial directions are compactified or not, and one expects that the same entropy formula 
should be derived after compactification.
This point was argued in \cite{4dim1} as follows.
In the corresponding supergravity solution, dilaton field diverges at the horizon.
Therefore this solution may receive large quantum correction and this provides the entropy 
expected from stringy calculation.

The situation must be the same in a black hole which is composed of D0-branes and 
intersecting D4-branes, related by U-duality to the compactified D1-D5 system 
mentioned above.
The corresponding supergravity solution indeed has zero area horizon and dilaton divergence 
at the horizon.
Because of U-duality invariance of entropy, it is expected that there exist microstates 
which provide the entropy.
Such microstates was discussed recently in \cite{Blum}, but their existence was not proved.

In this paper we will find out such microstates by constructing the low energy effective 
theory at the D4-D4 intersection.
In this theory D0-branes will appear as monopoles, just like the D4-D0 case in which 
D0-branes appear as instantons \cite{BinB}.
Then the entropy can be derived by counting the monopole ground states which preserve 
a half of supersymmetries.

This paper is organized as follows.
In section 2 we summarize the classical aspects of the black hole considered above.
In section 3 we construct the effective theory at the D4-D4 intersection.
Its monopole solutions are discussed in section 4, and the entropy formula is derived 
in section 5.
Section 6 is devoted to conclusion and discussion.
The properties of the monopole solutions are summarized in appendix.

\bigskip

{\large {\bf 2.\ Black hole solution and entropy}}

\medskip

The configuration of the D4-D0 system we will discuss is as follows. 
Consider Type IIA theory compactified on $T^6$.
There are $Q_1$ D4-branes wrapping, say, along (4567) directions of $T^6$, and $Q_2$ 
D4-branes wrapping along (6789) directions (we will denote this as D4'-branes). 
D4-branes and D4'-branes intersect at a 2-brane.
In addition, there are N D0-branes at the intersection. 
The corresponding solution of Type IIA supergravity is known \cite{solution}. 
It can be obtained from the solution of 11 dimensional supergravity by dimensional 
reduction \cite{reduction}. 

\newpage

\begin{eqnarray}
ds_{11}^2 &=& (H_1H_2)^{\frac23} [ (H_1H_2)^{-1}(-dt^2+dx_{10}^2+k(dt-dx_{10})^2)
              +H_2^{-1}(dx_9^2+dx_8^2) \nonumber \\
        & &  +(H_1H_2)^{-1}(dx_7^2+dx_8^2)+H_1^{-1}(dx_5^2+dx_4^2)+dx_3^2+dx_2^2+dx_1^2 ] 
\end{eqnarray}
\begin{eqnarray}
&& H_i = 1+\frac{c_iQ_i}r \hspace{5mm} (i=1,2) \hspace{4mm} , \hspace{4mm} 
     k = \frac{c_PN}r \nonumber \\
&& r^2 = x_1^2+x_2^2+x_3^2 \nonumber 
\end{eqnarray}
After dimensional reduction to 4 dimensions, this solution can be regarded as a black hole 
solution. 
Its metric in Einstein frame and dilaton field are
\begin{eqnarray}
ds_E^2           &=& -(1+k)^{-\frac12}(H_1H_2)^{-\frac12} dt^2
                     +(1+k)^\frac12(H_1H_2)^\frac12(dx_1^2+dx_2^2+dx_3^2) \nonumber \\
e^{-2\phi_{(4)}} &=& (H_1H_2)^{-\frac16}(1+k)^\frac12 . \label{classical}
\end{eqnarray}
This solution has zero area horizon, and its Bekenstein-Hawking entropy is zero. 

However, it can be shown that the above brane configuration is related by U-duality to the 
D1-D5 system compactified on $S^1$. 
The method of counting microstates used in \cite{CM} does not depend on whether the spatial 
directions are compactified or not, and this suggests that the same entropy formula is 
valid when compactified on $S^1$. 
Because of U-duality invariance of entropy, it is natural to expect that the entropy of 
D4-D4'-D0 system is, at quantum level,
\begin{equation}
  S = 2\pi \sqrt{Q_1Q_2N} . 
\end{equation} 
This is implied by the classical solution (\ref{classical}). 
One can see that the dilaton field diverges at the horizon, 
indicating that entropy may receive large quantum corrections. 
This point was argued in \cite{4dim1} by the dimensional arguments. 
In the following sections, we will look for the corresponding microstates 
in D4-D4'-D0 system which contribute to the entropy.

\bigskip

{\large {\bf 3.\ Effective theory on the intersection}}

\medskip

Low energy effective theory on the D4-D4' intersection is (2+1) dimensional ${\cal N}=4$ 
super Yang-Mills theory with matters. 
Its matter contents can be derived from string perturbation theory. 
Consider an open string attached to the D4-branes.
The massless states which come from the string on D4-branes 
form an ${\cal N}=1$ vector multiplet in 10 dimensions. 
This is an ${\cal N}=4$ vector multiplet and adjoint hypermultiplet in 3 dimensions.
There are similar multiplets coming from D4'-branes.
The gauge group is therefore $U(Q_1)\times U(Q_2)$. 
The massless states which come from the two strings stretching between D4 and D4' form an 
${\cal N}=4$ hypermultiplet in the representation $({\bf Q_1,\bar{Q}_2})$ 
for the gauge group. 

The action is almost entirely determined by supersymmetry. 
In this paper we will consider the simplest case, $Q_1=Q_2=1$. 
In this case, adjoint hypermultiplets decouple from the dynamics. 
Moreover, one of a linear combination of the vector multiplets also decouple and 
the effective theory is just ${\cal N}=4$ super QED with one fundamental hypermultiplet. 
It is convenient to obtain the action from 4 dimensional ${\cal N}=2$ theory by dimensional 
reduction. 

The bosonic part of the action is

\newpage

\begin{eqnarray}
S_{boson} &=& \int d^3x [\frac1{g^2}(-\frac14 F_{ij}F^{ij}
             -\frac12 \partial_i\phi_m\partial^i\phi_m)
             -D_iq^\dagger D^iq-D_i\tilde{q}^\dagger D^i\tilde{q} \nonumber \\
          & & \hspace{1cm} -\frac{g^2}2(q^\dagger q+\tilde{q}^\dagger \tilde{q})^2 
             -\phi_m\phi_m(q^\dagger q+\tilde{q}^\dagger \tilde{q})] \\
          & &D_iq=(\partial_i-iA_i)q, \hspace{5mm} 
             D_i\tilde{q}=(\partial_i+iA_i)\tilde{q} \nonumber \\
          & &(i,j=0,1,2 \hspace{3mm} m=1,2,3) , \nonumber 
\end{eqnarray}
where $\phi_m$ are the real scalars in the vector multiplet and $q,\tilde{q}$ are the 
complex scalars in the hypermultiplet. 
Note that $A_i=A_i^{(1)}-A_i^{(2)}$ where $A_i^{(1)}$ is a gauge field 
coming from the D4-brane and $A_i^{(2)}$ is from the D4'-brane.

The effective theory should be parity invariant.
The parity transformation exchanges $x_1$ and $x_2$, and reverses orientation of both 
D4-brane and D4'-brane.
But physically this makes no difference.
Therefore we will not include in the action parity-violating terms such as Chern-Simons term.
 
${\cal N}=4$ supersymmetry transformations of fermions are the following.
\begin{eqnarray}
\delta\lambda &=& (\frac12\gamma^{ij}F_{ij}+ig^2(q^\dagger q-\tilde{q}^\dagger\tilde{q}))\xi
                 +\gamma^i\xi\partial_i\phi_3-\sqrt2\gamma^i\eta^*\partial_i\phi
                 +2g^2\eta q\tilde{q} \nonumber \\
\delta\psi    &=& (\frac12\gamma^{ij}F_{ij}-ig^2(q^\dagger q-\tilde{q}^\dagger\tilde{q}))\eta
                 +\gamma^i\eta\partial_i\phi_3+\sqrt2\gamma^i\xi^*\partial_i\phi
                 -2g^2\xi q^\dagger\tilde{q}^\dagger \nonumber \\
\delta\psi_q  &=& \sqrt2\gamma^i\xi^*D_iq+\sqrt2i\gamma^i\eta^*D_i\tilde{q}^\dagger
                 -2\xi\tilde{q}^\dagger\phi^\dagger-2i\eta q\phi^\dagger
                 -\sqrt2i\xi^*\phi_3q+\sqrt2\eta^*\phi_3\tilde{q}^\dagger \nonumber \\
\delta\psi_{\tilde{q}}
              &=& \sqrt2\gamma^i\xi^*D_i\tilde{q}-\sqrt2i\gamma^i\eta^*D_iq^\dagger
                 -2\xi q^\dagger\phi^\dagger+2i\eta\tilde{q}\phi^\dagger
                 +\sqrt2i\xi^*\phi_3\tilde{q}+\sqrt2\eta^*\phi_3q^\dagger \label{susy}
\end{eqnarray}
$\lambda,\psi$ are the fermions in the vector multiplet, $\psi_q, \psi_{\tilde{q}}$ are in 
the hypermultiplet and $\phi=\frac1{\sqrt2}(\phi_1+i\phi_2)$. $\xi,\eta$ are 2-component 
Dirac spinors in 3 dimensions. $\gamma^i$ are 3 dimensional Dirac matrices. 

\bigskip

{\large {\bf 4.\ BPS monopole solutions}}

\medskip

In the effective theory constructed above D0-branes will appear as BPS monopoles preserving 
half of supersymmetries, 
as they appear in D4-brane effective theory as instantons \cite{BinB}. 
From supersymmetry transformations (\ref{susy}), one can show that the static solutions 
which preserve half of supersymmetries fall into two cases.
The first one is
\begin{eqnarray}
&& F_{12}=\epsilon g^2|q|^2,\hspace{5mm} D_1q-i\epsilon D_2q=0  \label{monoeq1} \\
&& \hspace{1cm} \tilde{q}=\phi_m=0 \nonumber
\end{eqnarray}
when the Grassmann parameters satisfy
\begin{equation}
(\gamma^0+\epsilon)\xi=(\gamma^0-\epsilon)\eta=0 .
\end{equation}
The second one is
\begin{eqnarray}
&& F_{12}=\epsilon g^2|\tilde{q}|^2, \hspace{5mm} 
   D_1\tilde{q}+i\epsilon D_2\tilde{q}=0 \label{monoeq2} \\
&& \hspace{1cm} q=\phi_m=0 \nonumber
\end{eqnarray}
when
\begin{equation}
(\gamma^0-\epsilon)\xi=(\gamma^0+\epsilon)\eta=0 ,
\end{equation}
where $\epsilon=\pm 1$.

Unfortunately, it is shown that these equations have no physically acceptable solution 
\cite{monopole}.
Now we perform an analytic continuation such as
\begin{equation}
x^a \to -ix^a, A_a \to iA_a \hspace{5mm} (a=1,2) .
\end{equation}
Then the equations (\ref{monoeq1})(\ref{monoeq2}) become
\begin{equation}
F_{12}=\epsilon g^2|q|^2, \hspace{5mm} D_1q+i\epsilon D_2q=0 \label{modif1}
\end{equation}
or,
\begin{equation}
F_{12}=\epsilon g^2|\tilde{q}|^2, \hspace{5mm} 
D_1\tilde{q}-i\epsilon D_2\tilde{q}=0. \label{modif2}
\end{equation}
The monopole solutions of (\ref{modif1})(\ref{modif2}) have been constructed explicitly in 
\cite{monopole} and we summarize these properties in appendix.
We will discuss $\epsilon=+1$ case below. 
$\epsilon=-1$ case corresponds to the anti D0-branes at the intersection.

The one-monopole solution to eqs.(\ref{modif1})(\ref{modif2}) is characterized by a positive 
integer \linebreak (monopole charge) and 3 real parameters (its position and scale). 
When we consider the superposition of monopoles, one more real parameter is needed 
for each monopole to parametrize the general solutions. 
Thus we conclude that the one-monopole solution has 4 real parameters.
Its moduli space is therefore ${\bf R}^2 \times {\bf C}$.
In the situation we are discussing in relation to the black hole, the topology of the
intersection is $T^2$.
Then the moduli space in this case will be, after suitable compactification,
\begin{equation}
{\cal M}_1=T^2 \times S^2
\end{equation}

The dynamics of the monopole is described by a supersymmetric quantum mechanics 
whose target space is ${\cal M}_1$.
Its ground states correspond to the cohomology classes of ${\cal M}_1$.
The number of ground states can be determined as follows.
\begin{eqnarray}
\mbox{(number of bosonic states)} &=& \sum_{k=0}^2 \mbox{dim}(H^{2k}({\cal M}_1))=4 \\
\mbox{(number of fermionic states)} &=& \sum_{k=0}^1 \mbox{dim}(H^{2k+1}({\cal M}_1))=4
\end{eqnarray}

\bigskip

{\large {\bf 5.\ Ground state counting and entropy}}

\medskip

Now we can derive the entropy formula by counting the ground states of monopoles.
Consider the general case $(Q_1,Q_2>1)$.
Each D4-brane is supposed to be separated from the other D4-branes (the same is true for 
the D4'-branes), and there are $Q_1Q_2$ D4-D4' intersections.
Assume that the dominant contribution to the entropy comes from the monopole states 
described in section 4 which are at the intersections.
Total monopole charge of these states correspond to D0-brane charge $N$.
Recalling that each monopole has 4 bosonic ground states and 4 fermionic ones, 
one can see that counting these states is equivalent to counting the degeneracy of states 
at level $N$ in free CFT of $4Q_1Q_2$ bosons and $4Q_1Q_2$ fermions.

The equivalence between the partitions of D0-branes into their ground states and 
the states in the CFT can be shown by the following one-to-one correspondence.
The partition of D0-branes is characterized by a set of integers,
\begin{eqnarray}
&& \{ n_B(k,i,Q),n_F(k,i,Q)\}, \\
&& k=1,2,\cdots,\hspace{3mm}i=1,2,3,4,\hspace{3mm}Q=1,2,\cdots ,Q_1Q_2 \nonumber \\
&& \sum_{k=1}^{\infty} \sum_{i=1}^4 \sum_{Q=1}^{Q_1Q_2} (kn_B(k,i,Q)+kn_F(k,i,Q))=N \nonumber
\end{eqnarray}
where $n_B(k,i,Q)$ is the occupation number of the $i$-th bosonic state of the monopole with 
magnetic charge $k$ at the $Q$-th intersection and 
$n_F(k,i,Q)$ is the one of the fermionic state.
This corresponds to the following state.
\begin{eqnarray}
&& \prod_{k=1}^{\infty} \prod_{i=1}^4 \prod_{Q=1}^{Q_1Q_5} (\alpha^{(i,Q)}_{-k})^{n_B(k,i,Q)} 
(\psi^{(i,Q)}_{-k})^{n_F(k,i,Q)}|0> \\
&& \sum_{k=1}^{\infty} \sum_{i=1}^4 \sum_{Q=1}^{Q_1Q_2} (kn_B(k,i,Q)+kn_F(k,i,Q))=N \nonumber
\end{eqnarray}
Thus the generating function of the number of such partitions coincides to the partition 
function of the free CFT.

\begin{equation}
Z = \prod_{n=1}^\infty \frac{(1+q^n)^{4Q_1Q_2}}{(1-q^n)^{4Q_1Q_2}} 
  = \sum_{n=0}^\infty d(n)q^n \label{partition}
\end{equation}

This formula is rederived from the consideration of the cohomology of multi-monopole moduli 
space.
First we consider only one intersection.
Since the monopole states are BPS states, the multi-monopole moduli space will be a symmetric 
product space.
\begin{equation}
{\cal M_N}=({\cal M}_1)^N/S_N \equiv S^N{\cal M}_1
\end{equation}
The dimension of the cohomology of such space can be determined from the following 
formula \cite{cohomology},
\begin{equation}
\sum_{n=0}^{\infty} \sum_{k=0}^{nd} (-1)^ky^kb_k(S^nX)q^n 
= \prod_{n=1}^{\infty} \prod_{k=0}^d (1-y^{k+(n-1)\frac d2}q^n)^{-(-1)^kb_k(X)},
\end{equation}
where $d=\mbox{dim}X$ and $b_k(X)$ is Betti number of $X$.
Let $X={\cal M}_1$ and $y=-1$, we obtain,

\newpage

\begin{eqnarray}
\sum_{n=0}^{\infty} \left(\sum_{k=0}^{4n} b_k({\cal M}_n)\right)q^n
&=& \prod_{n=1}^{\infty} \prod_{k=0}^4 (1-(-1)^kq^n)^{-(-1)^kb_k({\cal M}_1)} \nonumber \\
&=& \prod_{n=1}^\infty \frac{(1+q^n)^4}{(1-q^n)^4} \nonumber \\
&=& \sum_{n=0}^{\infty} d'(n)q^n \label{single}
\end{eqnarray}
Thus we conclude that the number of ground states of N D0-branes is
\begin{equation}
\sum_{k=0}^{4N} \mbox{dim}(H^k({\cal M}_N)) = d'(N)
\end{equation}
Then let us return to the general $Q_1,Q_2$ case.
The generating function of the number of ground states is just a product of 
eq.(\ref{single}).
\begin{equation}
Z=\left( \prod_{n=1}^\infty \frac{(1+q^n)^4}{(1-q^n)^4} \right)^{Q_1Q_2}
\end{equation}
This coincides to eq.(\ref{partition}).

The entropy of this system is logarithm of $d(N)$, and thus
\begin{equation}
S = 2\pi\sqrt{Q_1Q_2N} \label{entropy}
\end{equation}
for large N.
Eq.(\ref{entropy}) coincides with the entropy derived from D1-D5 system by stringy 
calculation, and agrees with the U-duality invariance of the entropy.

As mentioned in section 2, from the classical geometry the entropy should be zero.
But this is not a contradiction.
In the monopole solution, the existence of the hypermultiplet scalars $q,\tilde{q}$ 
is very important.
They come from an open string stretching between intersecting D4-D4'.
The appearance of this string is regarded as the resolution of the classical singularity.
 
\bigskip

{\large {\bf 6.\ Conclusion and discussion}}

\medskip

We have discussed entropy of the 4 dimensional black hole composed of D0-branes and 
the intersecting D4-branes.
From the classical geometry, this black hole has zero area horizon, and its entropy seems to 
be zero.
We constructed low energy effective theory at the D4-D4' intersection and found D0-brane 
bound states which appeared as monopoles.
By the explicit monopole solutions, it is shown that the monopole has 4 real parameters 
and its moduli space is $T^2 \times S^2$.
Then we have concluded that the monopole has 4 bosonic ground states and 4 fermionic ones.
From these results, we have derived the entropy formula.
This agrees with the formula for D1-D5 system, which is U-dual of our black hole.

The emergence of the entropy is expected already from the classical black hole solution.
The dilaton diverges at the horizon, signaling the large quantum correction.
The resolution of the singularity is due to the existence of the hypermultiplet scalars 
$q,\tilde{q}$, which comes from open string stretching between the intersecting D4-branes.

One subtle point in our argument is the necessity of analytic continuation.
Without this, there is no state preserving a half of supersymmetries.
A similar problem exists in D4-D0 bound states.
It is well known that U(1) gauge field, which corresponds to one D4-brane case, cannot have 
instanton solution.
The monopole solution discussed in this paper may be related to this situation and, 
therefore, the problem will not exist when $Q_1,Q_2>1$.
It is interesting to extend our argument to the non-Abelian case.

\bigskip

{\large {\bf Acknowledgments}}

\medskip

I would like to thank H.Itoyama, K.Murakami, A.Tsuchiya, T.Yokono for valuable discussions.
This work is supported in part by JSPS Research Fellowships.
\newpage

{\large {\bf Appendix: Monopole solutions}}

\medskip

In this appendix, we will discuss the following equations.
\begin{eqnarray}
F_{12}=g^2|q|^2 \label{eq1} \\
D_1q+iD_2q=0   \label{eq2}
\end{eqnarray}
The solutions were constructed explicitly in \cite{monopole}.
Here we will solve (\ref{eq1})(\ref{eq2}) and summarize the properties of the solutions.

(\ref{eq1})(\ref{eq2}) can be solved as follows.
Since $q$ is a complex scalar, it can be written as
\begin{equation}
 q=\rho^{\frac12}e^{i\omega} , \label{q}
\end{equation}
where $\rho$ and $\omega$ are real functions.
Substituting (\ref{q}) into (\ref{eq2}), gauge fields $A_a$ are shown to be written as
\begin{equation}
A_a = \frac12\varepsilon_{ab}\partial_b\log\rho+\partial_a\omega , \hspace{3mm} 
       (a,b=1,2) 
\end{equation}
where $\varepsilon_{ab}$ is antisymmetric and $\varepsilon_{12}=+1$.
From (\ref{eq1}), one obtains the Liouville equation for $\rho$.
\begin{equation}
\Delta\log\rho=-2g^2\rho \label{liouville}
\end{equation}
$\omega$ will be determined by requiring regularity of the solution.

First we consider the radially symmetric solutions.
The general solutions are known.
\begin{equation}
\rho(r)=\frac{4n^2}{g^2r^2}((\frac{r_0}r)^n+(\frac r{r_0})^n)^{-2} \label{soln}
\end{equation}
$n,r_0$ are integration constants.
Regularity at origin and infinity requires $n\geq 1$.
To avoid singularity in $A_a$ at origin, $\omega$ should be $(n-1)\theta$.
Then $n$ must be an integer to make $q$ single-valued.
Thus we conclude that the radially symmetric solutions are
\begin{eqnarray}
q &=& \frac{2n}{gr}((\frac{r_0}r)^n+(\frac r{r_0})^n)^{-1}e^{i(n-1)\theta} \nonumber \\
A_a &=& -2n\varepsilon_{ab}\frac{x_b}{r^2}\frac{(\frac r{r_0})^n}{(\frac{r_0}r)^n+
           (\frac r{r_0})^n} .
\end{eqnarray}

The monopole charge $m$ can be easily calculated.
\begin{equation}
m=\int d^2x F_{12} = g^2\int d^2x \rho = 4\pi n
\end{equation}
It is related to D0-brane charge $N$ as $N=\frac m{4\pi}$.

More general solutions of (\ref{liouville}) is also known.
\begin{equation}
\rho(r,\theta)=\frac4{g^2}\frac{|f'(z)|^2}{(1+|f(z)|^2)^2}
\end{equation}
$f(z)$ is a holomorphic function and $z=re^{i\theta}$.
The radially symmetric solution (\ref{soln}) corresponds to $f(z)=cz^{-n}$.
The multi-monopole solutions can be obtained by choosing $f(z)$ as
\begin{equation}
f(z)=\sum_{k=1}^{n}\frac{c_k}{z-z_k} ,
\end{equation}
where $c_k,z_k$ are complex parameters.
Total monopole charge of this solution is $m=4\pi n$.
$z_k$ is position of the k-th monopole and $c_k$ is related to its size.
Thus this solution depends on $4n$ real parameters, and this means that each monopole has 
4 real parameters.


\bigskip

\begin{thebibliography}{99}

\bibitem{SV}A.Strominger, C.Vafa, {\it Microscopic Origin of Bekenstein-Hawking Entropy}, 
Phys. Lett. {\bf B379} (1996) 99, hep-th/9601029.
\bibitem{CM}C.Callan, J.Maldacena, {\it D-brane Approach to Black Hole Quantum Mechanics}, 
Nucl. Phys. {\bf B472} (1996) 591, hep-th/9601029.
\bibitem{BH}J.Bekenstein, {\it Black Holes and Entropy}, Phys. Rev. {\bf D7} (1973) 2333; \\
S.Hawking, {\it Particle Creation by Black Holes}, Comm. Math. Phys. {\bf 43} (1975) 199. 
\bibitem{solution}M.Cvetic, D.Youm, {\it Dyonic BPS Saturated Black Holes of Heterotic 
String on a Six-Torus}, Phys. Rev. {\bf D53} (1996) 584, hep-th/9507090; \\
A.Tseytlin, {\it Extreme dyonic black holes in string theory}, Mod. Phys. Lett. {\bf A11} 
(1996) 689, hep-th/9601177; \\
I.Klebanov, A.Tseytlin, {\it Intersecting M-branes as Four-Dimensional 
Black Holes}, Nucl. Phys. {\bf B475} (1996) 179, hep-th/9604166.
\bibitem{4dim1}J.Maldacena, A.Strominger, {\it Statistical Entropy of Four-Dimensional 
Extremal Black Holes}, Phys. Rev. Lett. {\bf 77} (1996) 428, hep-th/9603060.
\bibitem{4dim2}C.Johnson, R.Khuri, R.Myers, {\it Entropy of 4D Extremal Black Holes}, 
Phys. Lett. {\bf B378} (1996) 78, hep-th/9603061. \\
G.Horowitz, D.Lowe, J.Maldacena, {\it Statistical Entropy of Nonextremal Four- Dimensional 
Black Holes and U-Duality}, Phys. Rev. Lett. {\bf 77} (1996) 430, hep-th/9603195.
\bibitem{Blum}J.Blum, {\it Supersymmetric Quantum Mechanical Description of Four
Dimensional Black Holes}, hep-th/9907101.
\bibitem{BinB}M.Douglas, {\it Branes within Branes}, hep-th/9512077.
\bibitem{reduction}J.Maharana, J.Schwarz, {\it Noncompact Symmetries in String Theory}, 
Nucl.Phys. {\bf B390} (1993) 3, hep-th/9207016; \\
A.Sen, {\it Electric Magnetic Duality in String Theory}, Nucl. Phys. {\bf B404} (1993) 109, 
hep-th/9207053.
\bibitem{monopole}R.Jackiw, S.Pi, {\it Self-Dual Chern-Simons Solitons}, Prog. Theor. Phys. 
Suppl. {\bf 107} (1992) 1.
\bibitem{lecture}J.Harvey, {\it Magnetiv Monopoles, Duality, and Supersymmetry}, 
hep-th/9603086.
\bibitem{cohomology}L.G$\ddot{\mbox{o}}$ttsche, W.Soergel, {\it Perverse sheaves and 
the cohomology of Hilbert schemes of smooth algebraic surfaces}, Math. Ann. {\bf 296} 
(1993) 235; \\
J.Cheah, {\it On the cohomology of Hilbert schemes of points}, J. Alg. Geom. {\bf 5} 
(1996) 479; \\
R.Dijkgraaf, {\it Fields, Strings, Matrices, and Symmetric Products}, hep-th/9912104

\end{thebibliography}
\end{document}